\documentclass{elsart3p}
\usepackage{graphicx}
\usepackage{amssymb}
\begin{document}

\begin{frontmatter}

\title{
Comparing the Dynamics of Skyrmions and Superconducting Vortices
} 
\author{
C.J. Olson Reichhardt$^1$, S.Z. Lin$^1$, D. Ray$^{1,2}$, 
and C. Reichhardt$^1$ 
} 
\address{
$^1$Theoretical Division,
Los Alamos National Laboratory, Los Alamos, New Mexico 87545, USA\\ 
$^2$Department of Physics, University of Notre Dame, Notre Dame,
Indiana 46556, USA
} 

\begin{abstract}
Vortices in type-II superconductors have attracted enormous 
attention as ideal systems in which to study nonequilibrium 
collective phenomena, since
the self-ordering of the vortices competes with quenched disorder and 
thermal effects.
Dynamic effects found in vortex systems include depinning, 
nonequilibrium phase transitions, creep, structural order-disorder
transitions,
and melting. 
Understanding vortex dynamics is also important for 
applications of superconductors which require the vortices either to remain
pinned or to move in a controlled fashion.
Recently, topological defects called skyrmions have been 
realized experimentally in chiral magnets. 
Here we highlight
similarities and differences between skyrmion dynamics and vortex
dynamics.
Many of the previous ideas 
and experimental setups that have been 
applied to superconducting vortices can also be used to study skyrmions.
We also discuss some of the differences between the two systems,
such as the potentially large contribution of the Magnus force 
in the skyrmion system that can dramatically alter 
the dynamics and transport properties.

\end{abstract}
\end{frontmatter}

\section{Introduction}
In a type-II superconductor in the presence of a magnetic field, 
flux enters the sample in the form of quantized 
units called  vortices. 
In most cases vortices interact with each other via 
repulsive interactions, leading to the formation of a 
triangular lattice which is the lowest
energy configuration \cite{1,2}. 
When a current is applied to the sample, 
a Lorentz force acts on the vortices and causes them to move in a direction
perpendicular to the applied current. 
Once the vortices are in motion, the material develops a finite resistance
and a finite voltage response.
In transport measurements, voltage-current curves can be used to study the
vortex motion.
In general, the sample contains some form of disorder that
creates regions in which the superconducting order parameter is suppressed.
These act as pinning sites for the vortices, 
holding the vortices immobile under an applied current up to
the critical current at which the vortices depin \cite{1,2}. 
For real-world applications, it is generally desirable to maximize
the critical current, 
and numerous studies have addressed how to enhance the
pinning in superconductors by 
various methods including
ion irradiation \cite{3} and 
the creation of tailored nanostructured pinning arrays \cite{5,6,7}.  
On the basic science level, vortices 
have proven to be a very rich system 
for studying collective dynamics in the presence of 
random or periodic substrates. The depinning
of vortices has been shown to exhibit properties of 
a nonequilibrium transition from a pinned disordered state 
to a dynamically fluctuating state \cite{randomreorg}, followed by
a transition to a dynamical moving crystal \cite{4} 
or moving smectic \cite{8,9,10}.
The vortex dynamics can be two-dimensional (2D), 
effectively 2D, or fully three-dimensional (3D), 
and the 3D systems can be anisotropic, such as 
in high temperature superconductors \cite{2,11,12}.
With various lithographic techniques
it is possible to create periodic pinning potentials 
with different types of symmetry to trap the vortices 
\cite{5,6,13,14,15,16,17,18,19,20}.  
In these 
structured pinning arrays, 
commensuration effects can occur 
when the number of vortices 
is an integer \cite{5,6,13} or rational fractional \cite{15,J} multiple 
of the number 
of pinning sites, and 
at the commensurate fields, different types of vortex 
crystal structures can  be stabilized \cite{6}.  
When the vortices
are driven over the periodic arrays, a rich 
variety of dynamical effects can arise such as soliton type flows, dynamical
ordering, and negative differential conductivity 
\cite{21,22,23,24}. 
If some form of asymmetry is added to the pinning arrays, 
it is possible to create vortex ratchet effects in which an ac driving force
induces a net dc motion of vortices
\cite{25,26,27,28}. 
There have also been proposals for novel flux logic devices where the vortices
act as the information carriers \cite{29}.                 

Recently another type of emergent particle called a skyrmion
has been discovered in condensed matter systems. 
Skyrmions were originally proposed as a model for 
baryons and mesons \cite{30}, and they were predicted 
to occur in condensed matter systems such as chiral magnets \cite{31}. 
In 2008,
triangular skyrmion lattices were observed via
neutron scattering in chiral magnets \cite{33}, 
and soon after, a number of
experiments using other types of imaging techniques 
revealed skyrmions in bulk and thin film samples \cite{34}. 
The skyrmions can be stabilized more easily in thin films. 
It was also shown that skyrmions can be driven with an applied
current which produces a Lorentz force on them,
and that there can be a finite
critical current required to set them in motion
\cite{35,36}. 
The small size of the skyrmion critical current 
has generated excitement since it could mean that skyrmions 
may have a significant advantage over magnetic domain walls in 
applications such as logic devices
\cite{35,M}.
Skyrmion depinning has been
studied in both continuum \cite{37,38} 
and particle-based simulations \cite{39}. 

Skyrmions appear to have many features in  common with vortices 
in that they both minimize their repulsive interactions by forming a
triangular lattice, 
can be driven with external currents, 
and exhibit a critical current for depinning.  
There are also several differences. 
For example, in skyrmions 
the Magnus term can be an important or even the dominant contribution 
to skyrmion motion, whereas in vortices
the Magnus term is generally negligible. 
The strong
effect of the Magnus force is conjectured to be responsible for
the low critical current in the skyrmion systems
\cite{37,38}.   
Here we outline some of the similarities and differences in the 
particle based dynamics of skyrmions and vortices and also discuss
how results found for vortex systems such as commensurations or ratchet 
effects could be realized in skyrmion systems.  

\section{Particle Based Simulations}   
Although vortices are extended objects with a core, in extreme type II
superconductors
they can 
be represented as pointlike or linelike objects, 
greatly facilitating numerical and theoretical studies of their
electrodynamic properties. 
In an effective 2D model of pointlike vortices,
a single vortex $i$ undergoes motion 
described by the following equation of motion:
\begin{equation}  
\eta \frac{d {\bf R}_{i}}{dt} = 
{\bf F}^{vv}_{i} +  {\bf F}^{P}_{i} +  {\bf F}^{D}_{i} +  {\bf F}^{T}_{i}. 
\end{equation} 
Here ${\bf R}_i$ is the location of vortex $i$,
$\eta=\phi_0^2d/2\pi\xi^2\rho_N$ is the damping constant, 
$d$ is the sample thickness, 
$\phi_0 = h/2e$ is the elementary flux quantum, 
$\xi$ is the superconducting coherence length,
and $\rho_{N}$ is the normal-state resistivity of the material.     
The vortex-vortex  interaction force is 
${\bf F}^{vv}_{i}=  \sum^{N_{v}}_{j\neq i}F_{0}K_1(R_{ij}/\lambda){\hat {\bf R}}_{ij}$, 
where $K_{1}$ is the modified Bessel function, 
$\lambda$ is the London penetration depth, 
$F_{0} = \phi_0^{2}/(2\pi\mu_{0}\lambda^{3})$,
$R_{ij} = |{\bf R}_{i} - {\bf R}_{j}|$ is the distance between 
vortex $i$ and vortex $j$, and the unit vector
${\hat {\bf R}}_{ij} = ({\bf R} - {\bf R}_{j})/R_{ij}$.    
This form of vortex-vortex interaction 
is appropriate for 3D superconductors 
in which the vortex acts as a rigid object along its length.
For thin film superconductors, the vortex-vortex interactions are 
much longer range and can be modeled with a logarithmic  
potential,  
${\bf F}_{i}^{vv}=-\sum_{j\ne i}^{N_v}{\nabla} V(R_{ij})$ 
where $V(R_{ij})\propto \ln(R/\lambda)$ \cite{J}. 
The force ${\bf F}^{P}$ from the pinning sites can be treated with
various models; here, we consider
parabolic attractive sites with
\begin{equation}
{\bf F}^{P}_{i} = -\sum^{N_{p}}_{k=1}(F_{p}/r_{p})({\bf R}_{i} - {\bf R}^{(p)}_{k})
\Theta[(r_{p} - |{\bf R}_{i} - {\bf R}^{(p)}_{i}|)/\lambda].  
\end{equation}
where ${\bf R}^{(p)}_{k}$ is the location of 
pinning site $k$, $F_{p}$ is the maximum pinning force, 
$r_p$ is the pin radius,
and $\Theta$ is the Heaviside 
step function.  
Langevin thermal kicks are applied using
the term ${\bf F}^{T}_{i}$ which has the following properties: 
$\langle {\bf F}^{T}_{i}\rangle = 0$ and
$\langle F_{i}^{T}(t)F^{T}_{j}(t^{\prime})\rangle 
= 2\eta k_{B}T\delta_{ij}\delta(t- t^{\prime})$, 
where $k_{B}$ is the Boltzmann constant. 
The driving force 
is from an externally applied current ${\bf J}$
resulting in a  Lorentz force ${\bf J} \times {\bf B}$ 
perpendicular to the applied current. 
It is also possible
to use flux gradient driven simulations in which the
vortices are added at one side of the sample and, through their own 
mutual repulsion, drive themselves
though the sample \cite{40}.  
When the vortices are
pinned in the absence of creep, then even for a finite current the system 
is dissipationless and the voltage response is zero. 
Once the vortices are in motion,
a finite voltage response develops that is 
proportional to the number of moving vortices 
and their velocity.
If the external drive is applied in the
$x$ direction, the following 
quantity is proportional to the voltage: 
$\langle V_{x}\rangle = \sum^{N_{v}}_{i= 1}{\bf v}_{i}\cdot {\hat {\bf x}}$, 
where ${\bf v}_{i} = d{\bf R}_{i}/dt$.
 
Additional dynamical terms such as a Magnus force can arise due to the 
interactions of the
external current with currents circling around the vortices. 
This reduces the net current 
on one side of the vortex and enhances it on the other side.
In most cases, for superconducting vortices the 
damping term is dominant so that the Magnus term is negligible \cite{2}.        

The initial skyrmion simulations employed continuum models utilizing the 
Landau-Lifshitz-Gilbert equation
\cite{37,38}, and showed that in the absence of disorder the skyrmions 
form a triangular 
lattice. 
Under an applied drive, the velocity versus applied force
curves 
indicate that there is a depinning transition \cite{37} 
as well as other phenomena at  higher drives where the skyrmion
lattice transitions into a chiral liquid state \cite{38}. 
More recently, a particle-based model for skyrmions was proposed \cite{39} 
inspired by Thiele's approach \cite{N}.  
Here, the
equation of motion is  
\begin{equation}  
\alpha{\bf v}_{i} = {\bf F}_i^{ss} + {\bf F}^{P}_{i} +
{\bf F}^{M}_{i} + {\bf F}^{L}_{i} + {\bf F}^T_{i},
\end{equation}
where ${\bf v}_{i}$ is the skyrmion velocity. 
The damping term $\alpha$ is due 
both to spin precession and conduction electrons localized in the core. 
The Lorentz force
${\bf F}^{L} = 2\pi\hbar e^{-1} {\bf \hat z} \times {\bf J}$ 
arises due to the 
emergent quantized flux that occurs in the presence of a finite current.
This term corresponds to the driving term in the vortex equation of motion. 
${\bf F}^{P}$ is the force from the pinning sites and
the Magnus term is ${\bf F}^{M}_{i} = \beta{\bf \hat z} \times {\bf v}_{i}$.
By conducting continuum simulations where one skyrmion is brought
close to a second immobilized skyrmion,
we showed \cite{39} that
the skyrmion-skyrmion interaction force ${\bf F}^{ss}$ 
can be well approximated by
$K_{1}(r_{d}/\xi)$, where $r_{d}$ is the distance between 
the skyrmions and 
$\xi$ is the size of the core. 
Using typical parameters from MnSi, 
an estimation of the
dissipative force per unit length is 
$5\times 10^{-6}$ N/m while the Magnus force is $5\times 10^{-5}$ N/m, 
indicating that the
Magnus force will play an important role.  

\section{Structure and Dynamics}

\begin{figure}
\begin{center}
\includegraphics[width=\columnwidth]{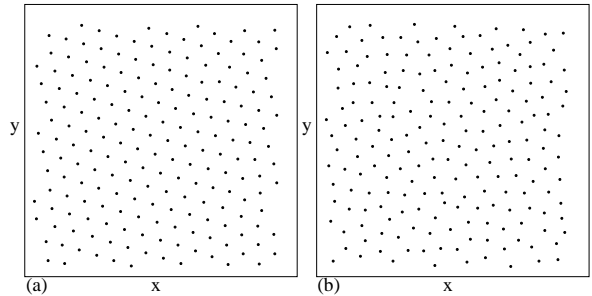}
\end{center}
\caption{Skyrmion positions (black dots) after 
simulated annealing. (a) A system without quenched disorder
where the skyrmions form a triangular lattice. (b)
A system with strong disorder shows a disordered or glassy structure.
}
\label{fig:1}
\end{figure}

For skyrmion dynamics, the relevant ratio is $\beta/\alpha$, 
where $\beta$ is the
coefficient in front of the Magnus term and $\alpha$ is the damping term. In 
real experimental systems, this ratio can vary from  $10$ to $40$.  
In Fig.~1(a) we show the skyrmion structure in the absence 
of disorder for a system
containing $N_{s} = 194$ skyrmions with 
$\beta/\alpha = 10.0$. The structure is obtained by slowly annealing
from a high temperature state and cooling to $T = 0$. 
Here a triangular lattice appears similar to the triangular 
vortex lattice found in superconducting systems.    
In Fig.~1(b) we plot the skyrmion positions for the same system but with 
$N_{p} = 100$ non-overlapping pinning sites with maximum 
force  $F_{p} = 1.0$ and pinning radius $r_{p} = 0.35$. 
In this case a disordered   
structure or skyrmion glass phase occurs. 
For a vortex system with strong disorder, 
the vortex positions are also strongly disordered 
and the system forms a glasslike state \cite{2}. 
Further  work could focus on whether there is a 
well defined transition to a disordered state 
as function of disorder strength or temperature.     

\begin{figure}
\begin{center}
\includegraphics[width=\columnwidth]{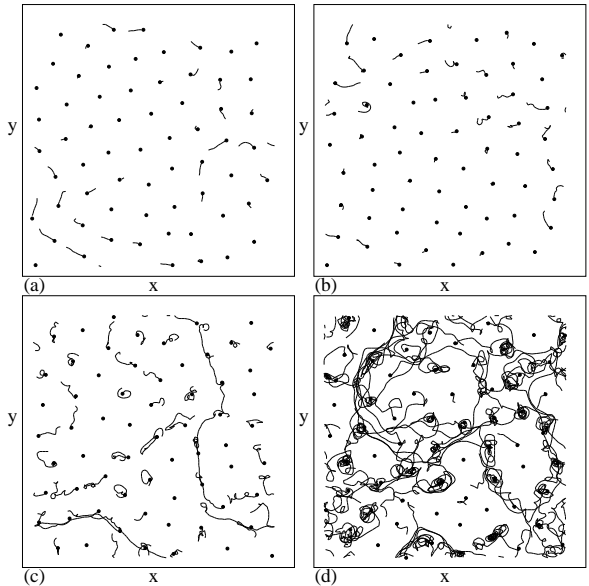}
\end{center}
\caption{ 
The locations (black dots) and trajectories (black lines) 
of the skyrmions in a system with $N_{s} = 64$ and $N_{p} = 51$
after a rapid thermal quench. 
The trajectories are obtained during a fixed period of time after the quench. 
(a) $\beta/\alpha = 0.0$. (b) $\beta/\alpha = 1.73$. 
(c) $\beta/\alpha = 10$. (d) $\beta/\alpha = 40$.
As the Magnus term $\beta$ increases, there is more motion after the 
quench with increasingly circular features. This
is consistent with previous work showing that increasing 
the Magnus term lowers the effectiveness of the pinning.  }
\label{fig:2}
\end{figure}

The role played by the Magnus term when the skyrmions are moving is revealed
 by conducting a thermal quench.
Here, we start the system 
in a higher temperature
state and rapidly cool to $T = 0$ in order to put the
system into a strongly nonequilibrium state. 
The particles undergo transient motion as they adjust and try to
lower their positional energy.   
Here we consider a system with $N_{s} = 64$, 
$N_{p} = 51$, and $F_{p} = 0.25$, and measure the particle trajectories
over a fixed period of time after the quench.
In Fig.~2(a) we show a system with $\beta/\alpha = 0$ which 
is the superconducting 
vortex case.  Here, 
a large number of particles are immediately immobilized by pinning sites,
and there are small patches of
transient motion where the particles move in almost straight paths.  
In this damping-dominated regime, the motion
quickly disappears.
In Fig.~2(b) at $\beta/\alpha = 1.73$, 
there are still a number of pinned particles;  
however, the trajectories of the moving particles are more circular.
For $\beta/\alpha=10.0$ in Fig.~2(c),
there are a larger number of particles in motion and  the
circular nature of the trajectories is more prominent. 
The increased motion indicates that the particles 
effectively experience a lower pinning force due to the Magnus term, which
has a tendency to cause particles to move perpendicular 
to an attractive force such as the pinning sites. 
This phenomenon 
was also observed
in continuum simulations and was argued to be the 
reason that pinning is weak in skyrmion systems \cite{37,39}. 
The system eventually settles into a motion free state after some time, 
and this transient time increases with increasing $\beta/\alpha$.
In Fig.~2(d), at $\beta/\alpha = 40$ the circular motion is much 
stronger and there are even fewer stationary particles.

\begin{figure}
\begin{center}
\includegraphics[width=\columnwidth]{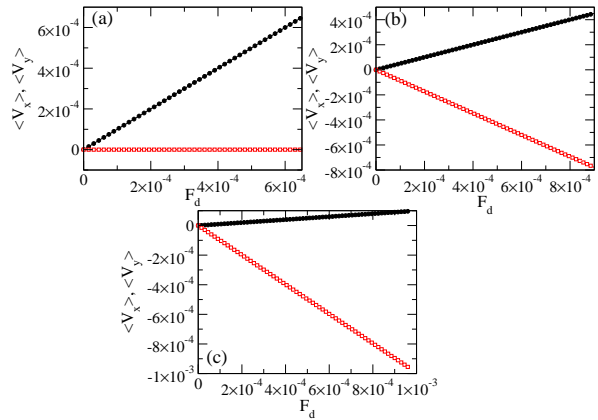}
\end{center}
\caption
{The average velocity per particle vs external drive $F_{d}$  
for a skyrmion system in the absence of quenched disorder. 
Filled circles (upper curves): 
$\langle V_{x}\rangle$;  open squares (lower curves): $\langle V_{y}\rangle$. 
(a) $\beta/\alpha = 0.0$, where 
$\langle V_{x}\rangle$ is finite and $\langle V_{y}\rangle = 0$. (b)
$\beta/\alpha = 3.87$, where the skyrmions move in both 
the $y$ and $x$-directions. (c) $\beta/\alpha = 10$, where the motion
is predominantly in the $y$-direction.}     
\label{fig:3}
\end{figure}

\section{Transport}

We next consider the transport for skyrmion systems in the absence of 
disorder. 
We apply a uniform Lorentz force 
of magnitude $F_d$ along the positive $x$ direction
to all the particles 
and measure the average velocity per particle vs $F_{d}$. 
In Fig.~3(a) we plot 
$\langle V_{x}\rangle$ and $\langle V_{y}\rangle$ versus 
$F_{d}$ for $\beta/\alpha = 0$. In this
case, the skyrmions form a lattice that moves along the $x$-direction with
a velocity parallel to the driving direction.
In Fig.~3(b), at 
$\beta/\alpha = 3.87$,
the skyrmions do not move strictly in the $x$ direction 
but move at angle of $-60^{\circ}$ with respect to the driving direction.
The angle of motion $\theta$ obeys $\tan(\theta)=\beta/\alpha$.
In Fig.~3(c) at $\beta/\alpha = 10$, 
the skyrmion motion is mostly in the $y$ direction at an angle 
of $-84^{\circ}$ to the driving direction. 
This result shows that as
the Magnus term increases, the skyrmions move increasingly in 
the transverse direction to 
an applied driving force. 
In an experiment, the driving force originates from
the Lorentz force from an applied current. 
This means that 
as the Magnus force increases, the skyrmions will 
move increasingly perpendicular to the Lorentz force and will come closer
to moving in the direction of the applied current,
in agreement with experiment \cite{36} and simulations \cite{37,38,39}.    
Future studies will focus on how the Magnus forces affects the 
transport curves in the presence of pinning sites.

\begin{figure}
\begin{center}
\includegraphics[width=0.8\columnwidth]{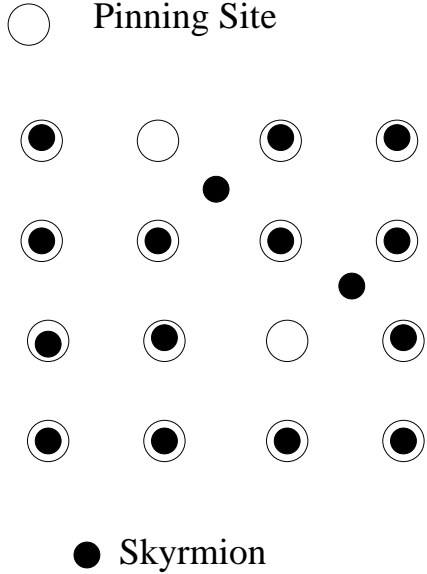}
\end{center}
\caption
{Schematic of skyrmions interacting with a periodic pinning
array. Here it may be possible to stabilize 
a skyrmion lattice with square symmetry
and to observe different types of commensuration 
effects similar to those found
in vortex systems with periodic pinning arrays.   
} 
\label{fig:4}
\end{figure}

\begin{figure}
\begin{center}
\includegraphics[width=\columnwidth]{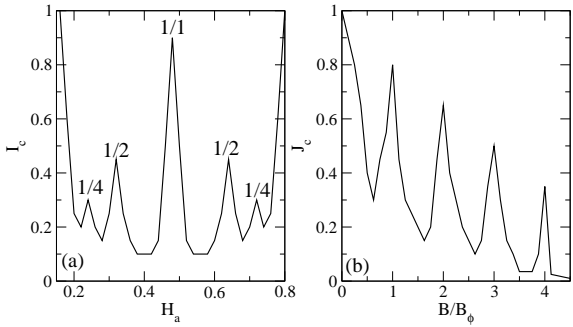}
\end{center}
\caption
{(a) Schematic of possible critical current behavior vs 
applied field $H_{a}$ for a skyrmion system in a square periodic pinning
array. In the skyrmion system, the density of skyrmions 
passes through a maximum at intermediate fields (shown in the
schematic as $H_a=0.5$). If the square periodic
pinning arrays has a one-to-one matching with the maximum
skyrmion density, then a peak in the critical current could occur 
at the one-to-one matching field, with smaller peaks
for 1/2 and 1/4 matching. The 1/2 and 1/4 matchings would occur twice.  
(b) A schematic of the behavior of the critical current $J_c$ of
superconducting vortices interacting with a square periodic pinning
array as a function of the applied magnetic field $B/B_\phi$, where
$B_\phi$ is the matching field.
Here, there are peaks at the integer matching fields; however, the critical
current generally decreases as the number of vortices increases.
}            
\label{fig:5}
\end{figure}

\section{Discussion}
It would be very interacting to explore how skyrmions would interact 
with artificial pinning arrays.  One
of the most basic questions is what types of commensuration 
effects would occur when the number of skyrmions
is an integer multiple of the number of pinning sites. 
It may also be possible to stabilize
new types of skyrmion lattice structures with square or other symmetries.
In Fig.~4 we show a schematic of skyrmions interacting with a
square periodic pinning array.  In the schematic it is 
assumed that the 
skyrmions are attracted to the pinning sites; however, 
it should be possible to create anti-pinning sites or areas that
repel the skyrmions by suitably modifying the exchange energy on a
local scale.
In the case of superconducting vortices, a common method for creating
artificial pinning sites is to use nanohole arrays
\cite{5,6}. 
It is not known if a skyrmion would be pinned by a nanohole or if it 
can even exist inside a nanohole.  
It should also be possible to create blind holes or regions where
the sample is thinner or thicker than average, although the effects of
such regions on skyrmions is unclear.
In vortex systems, the magnetic flux is real and is conserved;
however, for skyrmion systems, the flux carried by a skyrmion is
artificial and thus is not necessarily conserved,
so it is possible that thickness modulations would simply produce
different local skyrmion densities.
It is also possible that the size or shape of the
skyrmion could be modified by a pinning site.

If commensuration effects occur in skyrmion systems 
as a function of applied field, they would differ from those found
for superconducting vortices
since the skyrmion number is a non-monotonic function of the
applied magnetic field \cite{34}. 
As the field is increased from a low value,
the skyrmion number starts out
small and increases until reaching a maximum for an intermediate
field value.  As the magnetic field is further increased, the
skyrmion number decreases again before
reaching zero when the system enters a ferromagnetic  state. 
If a periodic pinning array were added to the sample
such that the number of pinning sites matches the peak 
number of skyrmions present, 
then the critical current as a function of 
applied field $H_{a}$ could have the form shown schematically in Fig.~5(a). 
The critical current is initially high for low skyrmion densities, and
passes through a commensurate peak at 
intermediate fields 
where there is a one-to-one matching of the skyrmions to the pinning sites.
It is also possible that commensurate effects 
would arise at rational fractional fillings such 
as $1/2$ and $1/4$, and due to the nonmonotonic dependence of skyrmion
density on field,
these matchings would occur twice as a function of
increasing field as shown in the schematic.
In Fig.~5(b)
this is contrasted with the
critical current behavior
for vortices in superconductors with periodic pinning sites,
where integer matching peaks appear at $B/B_{\phi} = n$, 
where $n$ is an integer and $B_\phi$ is the matching field. 
In the vortex system, the magnitude 
of the critical current generally decreases
with increasing $B$; in the skyrmion system, the critical current might
be highest when the density of skyrmions is lowest, which occurs at both
low and high values of magnetic field. 
It may also be possible to create asymmetric pinning geometries 
in order to realize 
skyrmion ratchets or diodes, similar to the ratchet 
effects observed in vortex systems with asymmetric pinning \cite{25,26,27,28}.
There is already some theoretical work showing how the Magnus term 
can lead to rectification effects \cite{41}.  

There is also the possibility that skyrmions could exhibit a 
melting transition for increasing temperature. In superconductors, there
is a phenomenon known as the peak effect that occurs 
as function of field or temperature which has been associated with a
softening of the vortex
lattice that allows the vortices to adjust their positions 
and become better pinned \cite{4}. 
It would  be interesting to see if a similar peak effect phenomenon
could occur in skyrmion  systems as the temperature is raised. 
Skyrmions are also expected to have tube like structures in 3D, so
a transformer geometry could be constructed where 
the top portion of the sample is driven while the response in the bottom 
portion of the sample is measured. 
In this way it could be possible to observe
cutting phenomena or entanglement effects 
or to measure the skyrmion line tension.

One difficulty with the experimental study of skyrmion 
motion compared to vortex 
systems is that for the vortex system, the motion is 
tied to the onset of dissipation. 
The skyrmion samples are not superconducting, 
so if a current is applied, the motion of the skyrmions can be masked by
a strong background dissipation signal, which must be separated from the
transport signal generated by the skyrmions.
This separation has already been achieved
in experiment; however, the skyrmion signal can be small \cite{35}. 
Another possibility would be to use imaging techniques
and directly measure the skyrmion motion, 
such as by using Lorentz microscopy \cite{36}. 

\section{Summary}
In summary, we have compared the particle based model of 
vortices in type-II superconductors to that of skyrmions in chiral magnets. 
The equations of motion for the two systems are similar; however, 
for the skyrmion systems the Magnus term is important and
can be  the dominant term influencing the dynamics of the skyrmion.
We show that in this model the skyrmions 
form a triangular lattice in the absence of pinning and a disordered 
state in the presence of strong pinning. 
The role of the Magnus force can be more clearly observed by conducting 
quenched simulations and observing the transient particle
motion in the presence of pinning. 
In the case where the Magnus term is dominant, the particles 
show a long time transient motion and are 
weakly pinned.    
Under an applied drive, the skyrmions move increasingly in the direction
transverse to the drive
as the contribution of the Magnus term
is increased.    
We also discuss how many of the ideas for pinning and dynamics 
of vortices could be applied to skyrmion systems, such as 
creating periodic pinning arrays,
ratchet geometries, or transformer geometries, 
or looking for peak effect type phenomena.    

This work was carried out under the auspices of the 
NNSA of the 
U.S. DoE
at 
LANL
under Contract No.
DE-AC52-06NA25396.

\end{document}